\documentstyle[preprint,eqsecnum,aps]{revtex}

 {\par\noindent{\underline{Proof} \quad}}{\hfill$\Box$\bigskip}
 {\par\noindent{\underline{Proof} of the theorem\quad}}{\hfill$\Box$\bigskip}
 {\par\smallskip\noindent{\underline{{\it Remark}} \quad}}{\par\smallskip}
 {\par\smallskip\noindent{\underline{{\it Fact}} \quad}}{\par\smallskip}
 {\par\smallskip\noindent{\underline{{\it Example}} \quad}}{\par\smallskip}
 {\par\smallskip\noindent{{\it Assumotion} \quad}}{\par\smallskip}
 {\par\smallskip\noindent{{\it Condition} \quad}}{\par\smallskip}

\begin{document}
\title{On the coherence verification of the continuous variable state in Fock space }
\author{Wang Xiangbin\thanks{email: wang@qci.jst.go.jp}, Matsumoto 
Keiji\thanks{email: keiji@qci.jst.go.jp}
   and Tomita Akihisa\thanks{email: a-tomita@az.jp.nec.com}
\\
        Imai Quantum Computation and Information project, ERATO, Japan Sci. and Tech. Corp.\\
Daini Hongo White Bldg. 201, 5-28-3, Hongo, Bunkyo, Tokyo 113-0033, Japan }

\maketitle
\begin{abstract} 
Almost all novel observable phenomena in quantum optics are related to the quantum coherence.
The coherence here is determined by the relative phase inside  a state.
Unfortunately, so far all the relevant experimental results in quantum optics 
are insensitive to the phase information of the coherent state. Lack of phase information may cause serious
consequences in many problems in quantum optics. For example, an ensemble of two mode squeezed
states is a classical ensemble if the phase in each state is totally random; but it is a 
non-classical ensemble if the phase in each state is fixed. 
As a timly application, verification of this type of phase information in
an ensemble of two mode squeezed states from the conventional laser is crucial to 
the validity of the continuous variavable quantum teleportation(CVQT) experiment.
Here we 
give a simple scheme to distinguish two different ensemble of states: 
the Rudolph-Sanders ensemble,
by which each squeezed states emitted has a uniform distribution from $0-2\pi$ on the phase value; 
and the van Enk-Fuchs ensemble, 
which emmits identical states with a fixed(but unknown) phase for every state. 
We believe our proposal can help to give a clear picture on whether the existing
two mode squeezed states so far are indeed non-classical states which can be used as the entanglement resource.

\end{abstract}
\newpage
Quantum coherence of different states plays a fundamentally important role in the whole subject
of quantum optics. The squeezed states have been widely used to demonstrate
various novel non-classical properties in the past.
In particular, a two mode squeezed state
\begin{eqnarray}
|r,\phi>=e^{re^{i\phi}(a_1a_2-a_1^\dagger a_2^\dagger)}|00>=\sum_{k=0}^\infty \lambda^ke^{ik\phi}|00\rangle,\label{pure}
\end{eqnarray}
where $a_i^{\dagger}$ and $a_i$ are bosonic creation and annihilation operators
respectively in the two mode Fock space, $|00>$ is the vaccum state, 
is  hopefully to be used as the entangled resource to carry out various novel tasks
in the creteria of quantum information\cite{chuang} such as the quantum teleportation\cite{ben1,pan,frusawa}.
 Recently, this state has been used as the enatnglement
resource to teleport a quantum state between two spatially separated parties\cite{frusawa}.
However, as it is pointed out by Rudolph and Sanders\cite{barry} that the
the phase information $\phi$ here has never been tested, 
therefore states as defined by
the above equation from certain source could have different phase $\phi$ for each different
 wavepackets. 
The lack of the 
phase information causes the loss of coherence of the state. 
 Although  lots of experiments have been done sucessfully
to reconstruct the continuous variable states in the past,  none of them is related
to the phase information. That is to say, a pure state is not the only possible state
that is compatible with the observed results. 
As it is noted\cite{barry} that, a conventional meassurement on optical fields, such as homodyne detection using lasers,
involving mixing of defferent incoherent fields and subsquent detection by energy absorption in 
photodetectors: all such meassurements are completely insensitive to any optical coherence. 
Actually, the state could be in arbitrary classical 
probabilistic distribution via the different phases. If we use the random  uniform 
distribution, the observed state is actually a mixed diagonal state in 
Fock space\cite{barry,molmer}, which is given by
\begin{eqnarray}
\frac{1}{2\pi}\int^{2\pi}_0|r,\phi><r,\phi|d\phi
=\sum^\infty_0 \lambda^{2n} |nn><nn|.\label{mix}
\end{eqnarray}  
Definitely, there is neither quantum coherence nor quantum entanglement in the above state.
The  state defined by eq(\ref{mix}) is a totally classical state. If the phase $\phi$ of
each states from certain source
are indeed uniformly distributed from zero to $2\pi$, 
the so called squeezed states will have little novel properties
in the practical use because in such a case all quantum coherence has been lost
and all the observable phenomena are just the same as that given by
classical optics. For example, we can never really take the advantage of
its squeezing property for certain quadrature variable( although a certain
quadrature operator ie indeed squeezed on a single wavepacket).

Due to the lack of the phase information, the observable quantum
coherence property have never been verified on an ensemble of squeezed states. 
To answer the question
 whether the conventional source can produce the nontrivil quantum coherent state 
 we must have a way to detect the phase information $\phi$. 
 
A timly application for this type detection is on the validity of the CVQT experiment\cite{frusawa}.
Recently, the phase information of the two mode squeezed states has drawn much attention of the physicists
due to the 
issue of quantum teleportation of the continuous variable state in Fock space\cite{frusawa}. 
Since the tomography
result is independent of the phase information $\phi$, it is also possible that
the the state used as the entanglement source in the quantum teleportation experiment
is a mixture of the states with different $\phi$, as defined by eq.(\ref{mix}). 
If this is the case, then no quantum state can be teleported by such a separable state.
Definitely, an ensemble of $N$ identical copies of pure states as defined by 
$|r,\phi><r,\phi|^{\otimes N}$ with the unknown $\phi$(but all states in the ensemble have the same 
$\phi$) 
is totally different from the ensemble as defined by eq.(\ref{mix}).
Unfortunetely, so far there has been no way to distinguish these two totally
different cases. 

Due to the unclearity of the phase $\phi$, there is a very hot discussions on the validity
of the entanglement resource used in the CVQT experiment\cite{barry,fuchs,fuchs1,barry1}. The discussions
can be sumarrized as the following:

1. Rudolph and Sanders: The phase $\phi$ in each of the squeezed states from a conventional 
laser source are uniformly 
distributed from zero to $2\pi$.
The correct form of the quantum state for the ensemble is given by eq(\ref{mix})   

2. van Enk and Fuchs: The traditional formalism is insufficient to describe the meachanism two mode 
squeezed states produced by the conventional laser source.
Using quantum de Finetti theorem one can see that the two mode squeezed states produced by a
conventional laser is essentially an ensemble of many coppies of identical two mode squeezed states
with a fixd $\phi$, though the value of $\phi$ is unknown. 

We may see that the validity of the CVQT experiment is now reduced to  which of the above
statements correctly describe the property of the  source which produses the two
mode squeezed states  in the CVQT experiment. That is to say, to know the validity of CVQT experiment,
we have to distinguish the Rudolph-Sanders source and the van Enk-Fuchs source.
In this letter we give a scheme to detect the phase information
$\phi$ in the two mode squeezed state.
That is to say, the meassurement result by our scheme is $sensitive$ to the quantum coherence.
Using our scheme, the ensemble of states defined by eq(\ref{pure}) or eq.(\ref{mix}) can be easily distinguished.
Obviously, the scheme has broad potential applications in the whole subject of quantum optics.
For example, given many copies of pure squeezed states, we can verify that they are indeed pure states with the same
phase $\phi$. For another example, given two Fuch sources, each source emmits many identical two mode squeezed
states, the phase of the state from each source $\phi_1$ and $\phi_2$ respectively, they are fixed and unknown,
our scheme can be  used detect the value $\phi_1-\phi_2$.
As an immediate application, the scheme can be used to judge whether the states used as the entangled
resource in a recent quantum teleportation\cite{frusawa} 
is the pure state as defined in eq.(\ref{pure}) or a mixed state defined by eq.(\ref{mix}). 
Consequently, this detection 
could give a a clear 
judgement on whether the experiment done in ref.\cite{frusawa} is essentially  
a quantum teleportation or a classical simulation of quantum teleportation.

Now we show how to distinguish a Roudolph-Sanders source and a van Enk-Fuchs source.
Lets first see what happens to the van Enk-Fuchs source, i.e.
 a photon source emmits the identical
pure squeezed states, all of them have the same but unknown phase $\phi$ . 
As it is noted in ref\cite{bow} that a two mode squeezed state can be produced with the specifically
chosen polarization for each modes as the following
\begin{eqnarray}
|\psi_1(\phi)>=\sum_{l=0}^{\infty} \lambda^l e^{il\phi}|l>_{ah}|l>_{bv},
\end{eqnarray} 
where the subscripts $a$, $b$ are for the mode $a$ and $b$ respectively, $h$ and $v$ represents 
the horizontal and vertical polarizations respectively.
Now we consider another wave packet from the same source, of which the   mode $a$ and $b$ is exchanged.
The quantum state for this wavepacket is  
$|\psi_2(\phi)>=\sum_{m=0}^{\infty}\lambda^m e^{im\phi}|m>_{av}|m>_{bh}$.
The total state is then given by
\begin{eqnarray}
|\Psi>=|\psi_1>|\psi_2>=\sum_{l=0}^{\infty}\sum_{m=0}^{\infty}\lambda^{l+m}e^{i(l+m)\phi}
|l>_{ah}|m>_{av}|l>_{bv}|m>_{bh}.
\end{eqnarray}
After taking a meassurement on the photon number of each mode, i.e., on the quantity $n=l+m$,
 the state $|\Psi>$ is collapsed to a 
specific entangled state. As it has been demonstrated in ref\cite{bow}, this type of meassurement
cab be carried out by either a quantum non-demolition meassurement\cite{duan} or 
by a more feasible way, the destructive photon counting with post selection. 
Suppose the meassurement result is $n$ for each mode, then the state
 after the meassurement is $\sum_{m=0}^{n}e^{i[(n-m)\phi+m\phi]}|n-m>_{ah}|m>_{av}|n-m>_{bv}|m>_{bh}$.
The phase information is clearly included in the state after the meassurement.
For simplicity, we consider only the cases of $n=1$ only. 
In the real experiment we can set an appropriate value of $\lambda$ so that we have a significant
probability to get the result of $n=1$. With such a setting,  given many coppies of state $|\Psi>$, 
we can always have 
a significant number of states with $n=l+m=1$ after the meassurement. 
In the case of $n=1$, the state is
\begin{eqnarray}
|\Psi_1>=e^{2i\phi}(|1>_{ah}|0>_{av}|1>_{bv}|0>_{bh})+|0>_{ah}|1>_{av}|0>_{bv}|1>_{bh}.
\end{eqnarray} 
There is only one photon in each mode for the satte defined above. Obviously, the state defined above
can be rewritten in the following way
\begin{eqnarray}
|\Psi_1>=e^{2i\phi}(|H>_a|V>_b+|V>_a|H>_b).
\end{eqnarray}  
Where the states $|H>$, $|V>$ are for a horizontal or a vertical polarized photon states respectively.
For this state, meassurement results for the polarization of the two modes are always  $different$.
Clearly, if the states given from the source are indeed pure states, i.e., they are
many coppies of $|\Psi_1>$, then we can rotate the polarizers by the same angle to both modes, 
and in principle we can find certain
angle($\pi/4$) by which the meassurement result of the polarization in the two modes 
are always $same$ !
On the other hand, if the given states from the source are  mixed states as defined in
eq(\ref{mix}), i.e., phase $\phi$ in each wave packet can be different, we will have a 
clearly different observation results.
In such a case, by the same operation, for all cases we get the meassurement of $l+m=1$,
 the state is $\frac{1}{ 2}(|HV><HV|+|VH><VH|)$. The correlation
between the two modes will linearly decreased with $\cos^2\beta$ when the 
polarizers are rotated, here $\beta$ is the angle that is rotated. Consequently, a van Enk-Fuchs 
source and a Rudolph-Sanders source  can be distinguished in the following way:

Initially, in both cases they are totally negatively correlated, i.e., whenever a 
photon is detected
in mode $a$, there most be no photon detected in mode $b$, and vice versa.
 We rotate the polarizers by $\pi/4$, the van Enk-Fuchs source(see eq(\ref{pure})) 
will give a totally positive correlation for the meassurement result, while
the Rudolph-Sanders source(see eq(\ref{mix})) will give no correlation for the meassurement. 
Also, in rotating the polarizers,
if the correlation does not decrease linearly with $\cos^2\beta$, the source must be not 
a Rudolph-Sanders one. 

Obviously, the above scheme can be also used to detect the phase difference for two van 
Enk-Fuchs sources.
Now $|\psi_1>$ and $|\psi_2>$ are collected from two different sources whose unknown fixed phase
are denoted by $\phi_1$ and $\phi_2$ respectively. After the meassurement on the photon number basis
 of each mode
is done, the state up to a global phase factor is 
\begin{eqnarray}
|\Psi_1>=|H>|V>+e^{i(\phi_2-\phi_1)}|V>|H>.
\end{eqnarray}  
We rotate the polarizers continuously and test the correlations at each stage. There must be
 certain angle $\beta_0$ at which we can observe that the meassurement results of the two modes 
are totally 
positively correlated. This angle $\beta_0$ determines the phase difference $\phi_2-\phi_1$.

Note that in a real experiment, we actually do not need to take a meassurement on the basis of
of the photon numbers of each mode, i.e. $\sum_{n=0}^\infty|n><n|$, where $n=l+m$. The meassure
on such a basis could be difficult by our current technology. 
The only subtle task for us
is to collect the wave packet $|\psi_1>$ and $|\psi_2>$ so that they are indistiuguishable.
Once we can make $|\psi_1>$ $|\psi_2>$ indistinguishable we can simply carry out the detection 
scheme  by the usual photon counting and obtain the correct result to a good approximation.
Suppose we use port A and port B to detect the photons from mode A and B respectively. A polarizer
is placed before each  port. The polarizers are placed horizontally  in the begining. 
We choose a very 
small $r$ so that $r^2<<1$ for the photon source. In the photon counting, we only record the 
data of those events where at least one port detects one photon $and$ 
no port detects more than one photon.
For simplicity, we name the events satisfying this condition as "good events". We will analyse
the correlation between the two modes only using the data of the good events. 
In the whole process, there is only a very small probability that
 a good event is caused by a state with $l+m\not=1$, i.e., 
the wave packet has been
collapsed to a state of which the photon number of each mode is not 1. It's easy to see that this
type of event can only happen with a small ptobability. First, $l+m=0$ is impossible, because 
whenever we observed a good event, we have observed one photon at least in one port, so the photon
number of that mode must not be 0, consequently $l+m\not=0$. 
Second, $l+m=2$ or a larger number is possible but the chance is
negligible. We have already set $r$ to be very small. Since $r^2<<1$, the probability that the state
$|\Psi>$ collapsed to a state with $l+m=1$ is much larger than that of a state with $l+m>1$. 
For example, taking $r=0.01$, in average, we can have one good event from 10000 events, 
and the probability
that this good event is caused by a stete with $l+m=1$ is several thousands times larger 
than that by a state with $l+m\not=1$. 
Actually, to a good approximation, we even do not have to require the photon detector to 
distinguish 1 photon case from many photon case
in the detection if $r$ is small enough. Because once the detector detects photons, the probability of
1 photon is much higher than other cases. For clarity, we insist on using the term "1 photon" in the 
rest parts of the letter. However, whenever the term "one photon" is used, it can be approximately
understood as "any number of photons".  

Initially, all good evnts must be the events on which 1 photon detected in certain mode and 0 photon is 
detected in another mode. If the source is the van Enk-Fuchs source, when the polarizers are rotated
by $\pi/4$, a photon in each mode must be detected to all good events. 
If the source is the Rudolph-Sanders
source, when the polarizers are rotated by $\pi/4$, to all good events, half of them are $(1_A,0_B)$
or $(0_A,1_B)$ and half of them are $(1_A,1_B)$, where the 0 or 1 represents the number of photons 
detected, subscripts $A,B$ represent the the port $A,B$(or the mode $a,b$) respectively. 

The phase difference can also be detected
for two Enk-Fuchs sources. In this case, we have to rotate the polarizers continuously. 
The phase difference
$\phi_2-\phi_1$ is determined by the angle $\beta_0$ we observed 1 photon in each mode to
all good events.  
Thus we see, to a good approximation,
our scheme can carried out by  the normal photon counting technique.    

{\bf Acknowledgement:} We thank Prof Imai for support. 

\end{document}